\documentclass[11pt]{article}
\usepackage{amsmath,amssymb,amsthm}
\usepackage{url}
\usepackage[a4paper,margin=1in]{geometry}
\newtheorem{theorem}{Theorem}
\newtheorem{corollary}{Corollary}
\newtheorem{observation}{Observation}

\title{It Pays to Brag? On Counterfactual Fair Credit}
\author{
	Ilan Doron-Arad\\
	{\small MIT}\\[-1pt]
	{}
	\and
	Elchanan Mossel\\
	{\small MIT}\\[-1pt]
	{}
}
 \begin{document}
 \maketitle

\begin{abstract}
    Intellectual work, such as software, legal documents, and scientific results, typically builds on the work of others, and the same end product can often be obtained in several different ways. When assigning credit for such work, it is not clear whether credit should go according to the way the work was actually produced or also consider counterfactual ways of obtaining it.
    We study this question by generalizing Shapley values to account for counterfactual ways of obtaining the same result. We show that the fairness axioms of efficiency, nullity, and submodularity (in particular,  additivity) are incompatible with incentivizing creators to maximize social welfare. In fact, fair credit can uniquely incentivize arbitrarily small welfare relative to the optimum. We also show that without submodularity, incentive compatibility is achievable, and that the four classical fairness axioms uniquely determine the Shapley value and prohibit dependence on counterfactual outcomes.
\end{abstract}

Intellectual work is rarely produced from scratch, and the same end product can usually be obtained in more than one way. A software engineer can ship a feature by building on an existing open-source library or by reimplementing the functionality in-house. A lawyer can draft a contract starting from a colleague's precedent or from first principles. A scientist can establish a result relying on one body of earlier work or on another. For a classical illustration of the last case, in 1735 Euler~\cite{Euler1740} derived the celebrated Basel identity $\sum_{n=1}^{\infty}n^{-2}=\pi^2/6$ using Newton’s identities, yet the same result was later given a simple geometric proof~\cite{Wastlund2010} based on the Pythagorean theorem.

This raises a question: When assigning credit to those who contributed to a piece of intellectual work, should we reward only those whose contributions were actually used (the authors of the library, the precedent, Newton), or also those whose contributions provide an alternative route to the same product (Pythagoras)?
Moreover, could fair credit incentivize the creator to choose the route that maximizes the social welfare, that is, the value that the community derives from the work (e.g., the most maintainable code, the clearest contract, the simplest proof)?
In other words, should credit follow the actual contribution, or also consider counterfactual ways of obtaining the result?
The answer is not a priori clear.

The seminal work of Shapley~\cite{Shapley1953} shows that there is a unique way of assigning credit satisfying four basic fairness axioms: efficiency, additivity, nullity, and symmetry. In this paper, we generalize the credit sharing model of Shapley to enable accounting for counterfactual credit. We show that fundamental fairness axioms for credit assignment are incompatible with incentivizing creators to maximize social welfare. In fact, fair credit can uniquely incentivize the worst possible outcome, with arbitrarily small welfare relative to the optimum, and cannot force the creator to choose the option that minimizes her marginal contribution.  
This holds even under weaker axioms than the ones used by Shapley~\cite{Shapley1953} when considering counterfactual alternatives. We discuss below relevant prior work; then define the model, and state and prove our results.

\subsection*{Related work}
Credit allocation rules generalized Shapley values~\cite{Shapley1953} to
random-order values~\cite{Weber1988}, precedence-constrained
values~\cite{FaigleKern1992}, and asymmetric and causal
variants~\cite{Frye2020,Heskes2020}. Scientific credit allocation has also been studied using citation patterns~\cite{ShenBarabasi2014}.
The notion of counterfactual credit is studied in reinforcement learning,
where alternative actions are used to define learning
signals~\cite{Foerster2018,Mesnard2021,Meulemans2023,Li2021,Li2026}.
In contrast, we ask whether credit can depend on counterfactual alternatives
while all coalition values of the realized result remain fixed.
This does not rule out using counterfactuals to define the game itself.
Other impossibility results in attribution address the ability
of feature attributions to explain model behavior~\cite{Bilodeau2024}. Other related work studies credit and incentives in research
teams~\cite{OzerturkYildirim2021,OzerturkYildirim2025},
credit for algorithmic progress~\cite{Kotthoff2018},
and causal attribution of changes in system
metrics~\cite{Sharma2022}.
None of these works studies how alternative ways of obtaining
the same result affect credit for the realized construction.

\subsection*{Model and Definitions}
Fix players $N=\{0,1,\ldots,n\}$, $n\ge1$, with $0$ the {\em creator} player,
and a finite, nonempty set $X$ of possible ways for player $0$ to obtain her result. Each $x\in X$ represents a different construction of the same result (e.g., a different implementation of one software feature, a different draft of one contract, or a different proof of one theorem).
Let $\mathcal V$ be the set of all normalized, monotone {\em value} functions
$v:2^N\times X\to\mathbb R_{\ge0}$, that is, functions satisfying
\[
\begin{array}{@{\bullet\;}l@{\quad}l@{\quad}r@{\;}c@{\;}l}
	(\forall x\in X) & &
	v(\varnothing,x) &=& 0,\\
	(\forall x\in X) & (\forall S\subseteq T\subseteq N) &
	v(S,x) &\le& v(T,x).
\end{array}
\]
The total value $v(N,x)$ may vary across constructions $x\in X$.
We call $v(N,x)$ the {\em social welfare} of outcome $x \in X$.
For credit in intellectual work, the above assumptions are essential as we think of the function $v$ as how much the contributions of the players in $S \subseteq N$ contributed to obtain $x \in X$. Indeed, $S = \emptyset$ does not contribute, any $S$ contributes a nonnegative value, and the contribution can only increase as the set $S$ increases.
 Denote addition of value functions by 
\[
(\forall v,w\in\mathcal V)\,(\forall S\subseteq N)\,(\forall x\in X)\qquad
(v+w)(S,x)=v(S,x)+w(S,x).
\]
A different function $w \in \mathcal{V}$ can be thought of as representing a different result, while $v+w$ represents their combined value.
We denote by $r \in X$ the {\em realized} construction by player $0$; the other elements $X \setminus \{r\}$
are {\em counterfactual} constructions.
Define the residual value that $i$ contributes on top of $S$ to $x$ by
\[
(\forall i\in N)\,(\forall x\in X)\,(\forall S\subseteq N\setminus\{i\})\qquad
\Delta_i v(S,x)=v(S\cup\{i\},x)-v(S,x).
\]

\paragraph{Axioms.}
A {\em credit rule} is $\phi:\mathcal V\times X\to\mathbb R_{\ge0}^{N}$.
Consider the following elementary fairness axioms that the credit rule must satisfy. The first axiom is called \emph{nullity}, which asserts that players that did not contributed anything under any counterfactual option do not get credit:
\begin{equation}\tag{N}\label{ax:N}
	(\forall v\in\mathcal V)(\forall r\in X)(\forall i\in N)\quad
	[(\forall x\in X)(\forall S\subseteq N\setminus\{i\})\,
	\Delta_i v(S,x)=0]\implies\phi_i(v,r)=0.
\end{equation}
This generalizes the nullity axiom for the standard setting of Shapley for $|X| = 1$. More basic fairness axioms  are
\emph{efficiency} and \emph{additivity}, meaning that all credit is distributed among the players and credit is additive in the value functions, respectively: 
\begin{alignat}{2}
	(\forall v\in\mathcal V)(\forall r\in X)
	&\quad \sum_{i\in N}\phi_i(v,r) &={}& v(N,r),
	\tag{E}\label{ax:E}\\
	(\forall v,w\in\mathcal V)(\forall r\in X)
	&\quad \phi(v+w,r) &={}& \phi(v,r)+\phi(w,r).
	\tag{A}\label{ax:A}
\end{alignat}
The first two axioms (i.e., (N) and (E)) are very natural in credit assignment: a player who did not contribute at all does not get credit and all credit is distributed. However, additivity is harder to defend as there may be some dependencies between the value functions. Thus, we consider a weaker form of additivity called submodularity, requiring that credit satisfies a diminishing return property as a function of the value functions. 
Let $v\vee w$ and $v\wedge w$ denote the
pointwise maximum and minimum, respectively:
\[
\begin{aligned}
(\forall v,w\in\mathcal V)\,(\forall S\subseteq N)\,(\forall x\in X)\qquad
(v\vee w)(S,x)&=\max\{v(S,x),w(S,x)\},\\
(\forall v,w\in\mathcal V)\,(\forall S\subseteq N)\,(\forall x\in X)\qquad(v\wedge w)(S,x)&=\min\{v(S,x),w(S,x)\}.
\end{aligned}
\]
Note that both operations are closed under $\mathcal V$.
Inequalities between credit vectors are componentwise. Now, consider the \emph{submodularity} axiom: 
\begin{equation}\tag{SM}\label{ax:SM}
(\forall v,w\in\mathcal V)\,(\forall r\in X)\qquad
\phi(v\vee w,r)+\phi(v\wedge w,r)
\le\phi(v,r)+\phi(w,r).
\end{equation}
This expresses diminishing returns in credit attribution: overlapping contributions
should not generate additional credit when considered together.
Additivity (A) is a special case of (SM): since
$(v\vee w)+(v\wedge w)=v+w$, (A) gives equality in (SM).

\paragraph{Incentive compatible credit functions.}
Henceforth, assume the history and the value function $v \in \mathcal{V}$ remain fixed, and player $0$ may select any
$r\in X$, with utility $\phi_0(v,r)$.
Call player $i\in N$ \emph{influential} for $v\in\mathcal V$ if it contributes to at least one configuration: 
\[
(\exists x\in X)\,(\exists S\subseteq N\setminus\{i\})\qquad \Delta_i v(S,x)>0.
\]
A credit rule $\phi$ is \emph{incentive compatible} (IC) if the incentive of player $0$ aligns with maximizing social welfare whenever she is influential:
\begin{equation}\tag{IC}\label{ax:IC}
(\forall v\in\mathcal V\text{ such that player $0$ is influential for }v)\quad
\arg\max_{r\in X}\phi_0(v,r)
\subseteq
\arg\max_{r\in X}v(N,r).
\end{equation}
Note that 
the influential condition is necessary because, by (N), a non-influential player receives zero credit for every construction, even when the constructions have different social values.

\paragraph{Counterfactual-dependent credit.}
Define the restriction of the value function to the realized (non counterfactual) outcome:
\[
(\forall v\in\mathcal V)\,(\forall r\in X)\,(\forall S\subseteq N)\,(\forall x\in X)\qquad
v_r(S,x)=v(S,x)\mathbf1_{\{x=r\}}.
\]
Observe that the functions $v_r$, for $v \in \mathcal{V}$ are exactly the set of normalized monotone functions in the standard model of Shapley~\cite{Shapley1953} with only one possible outcome $|X| = 1$. 
A credit rule $\phi$ is \emph{counterfactual-dependent} if
\[
(\exists v\in\mathcal V)\,(\exists r\in X)\qquad \phi(v,r)\ne\phi(v_r,r),
\]
and \emph{counterfactual-independent} otherwise, that is, if
\[
(\forall v\in\mathcal V)\,(\forall r\in X)\qquad \phi(v,r)=\phi(v_r,r).
\]

\subsection*{Results}
Our main result is that the fundamental fairness axioms of efficiency and nullity combined with submodularity prevent, in the counterfactual setting where $|X| \geq 2$, a credit assignment that respects the incentive compatibility property. As (A) is in particular (SM), the statement can also be stated with additivity instead of submodularity.
\begin{theorem}\label{thm:irrelevance}
Suppose $|X|\ge2$.
\begin{enumerate}
\item Every \textnormal{(IC)} credit rule satisfying (N) is counterfactual-dependent.
\item Every rule satisfying (E) and \textnormal{(SM)} is counterfactual-independent.
\end{enumerate}
\end{theorem}

Thus (E), (SM), and (N) preclude (IC). More strongly, our proof shows that the incentive of the player together with the fairness axioms can lead to an arbitrarily small fraction of the optimal social welfare. Formally,
For every $\varepsilon>0$, there exist $v\in\mathcal V$ for which
player $0$ is influential and $b\in X$ such that every credit rule
$\phi$ satisfying (E), (SM), and (N) satisfies $0<v(N,b)<\varepsilon\max_{r\in X}v(N,r)$ and  
\[
\begin{aligned}
	\arg\max_{r\in X}\phi_0(v,r)
	&=\arg\min_{r\in X}v(N,r)=\{b\},\\
	\arg\min_{r\in X}\phi_0(v,r)
	&=\arg\max_{r\in X}v(N,r)=X\setminus\{b\}.
\end{aligned}
\]

\begin{proof}
For (1), suppose a credit function $\phi$ satisfies (N) and is counterfactual-independent.
Fix $\varepsilon>0$, $b\in X$, and $0<\delta<\min\{1,\varepsilon\}$.
Define
\[
(\forall S\subseteq N)\,(\forall x\in X)\qquad
v(S,x)=\delta\mathbf1_{\{x=b,\,0\in S\}}+
\mathbf1_{\{x\ne b,\,1\in S\}}.
\]
For every $x\in X$, the function $v(\cdot,x)$ is normalized, nonnegative, and monotone, so $v\in\mathcal V$.
Also, $\Delta_0v(\varnothing,b)=\delta>0$, so player $0$ is influential. Recall that $v_r(S,x)=v(S,x)\mathbf1_{\{x=r\}}$.
Player $0$ is null in $v_r$ whenever $r\ne b$:
\[
(\forall r\in X\setminus\{b\})\,(\forall x\in X)\,(\forall S\subseteq N\setminus\{0\})\qquad
\Delta_0v_r(S,x)=0.
\]
Hence, (N) and counterfactual-independence give
\[
(\forall r\in X\setminus\{b\})\qquad
\phi_0(v,r)=\phi_0(v_r,r)=0.
\]
Since $\phi_0(v,b)\ge0$, we have that $b$ maximizes the credit of player $0$, although
\[
(\forall r\in X\setminus\{b\})\qquad v(N,b)=\delta<1=v(N,r).
\]
Thus, (IC) fails.
If (E) also holds, every player other than $0$ is null in $v_b$, that is,
\[
(\forall i\in N\setminus\{0\})\,(\forall x\in X)\,(\forall S\subseteq N\setminus\{i\})\qquad
\Delta_i v_b(S,x)=0,
\]
so (N) gives
$\phi_i(v_b,b)=0$ for every $i\in N\setminus\{0\}$. Therefore,
\[
\phi_0(v,b)=\phi_0(v_b,b)
=\sum_{i\in N}\phi_i(v_b,b)=v_b(N,b)=\delta.
\]
Thus, $b$ uniquely maximizes  the credit of player $0$ and uniquely minimizes welfare;
all other outcomes minimize her credit and maximize welfare. Moreover, $$0<v(N,b)=\delta<\varepsilon\max_{r\in X}v(N,r).$$

Next, we prove (2), which proves the stronger consequence under (E), (SM), and (N).
For (2), assume (E) and (SM).
Fix $v \in \mathcal{V}$ and $r \in X$.
Both $v_r$ and $v-v_r$ belong to $\mathcal V$: for every $x\in X$, the functions $v_r(\cdot,x)$ and
$(v-v_r)(\cdot,x)$ are either equal to $v(\cdot,x)$ and zero, respectively, when $x = r$, or zero and $v(\cdot,x)$ otherwise. Hence, both belong to
$\mathcal V$. Since $v_r(N,r)=v(N,r)$, we have
$$(v-v_r)(N,r)=v(N,r)-v_r(N,r)=0.$$
Thus, by efficiency and nonnegativity of the credit rule $\phi$ we have
\[
\sum_{i\in N}\phi_i(v-v_r,r)=(v-v_r)(N,r) = 0.
\]
This implies
\[
(\forall i\in N)\qquad \phi_i(v-v_r,r)=0.
\]
Moreover, at most one of $v_r(S,x)$ and $(v-v_r)(S,x)$ is nonzero:
\[
(\forall S\subseteq N)\,(\forall x\in X)\qquad
v_r(S,x)=0\ \text{ or }\ (v-v_r)(S,x)=0.
\]
Thus,
$v_r\vee(v-v_r)=v$ and $v_r\wedge(v-v_r)=0$.
Applying (SM) coordinatewise, we obtain
\[
\begin{aligned}
(\forall i\in N)\qquad
\phi_i(v,r)+\phi_i(0,r)
&\le \phi_i(v_r,r)+\phi_i(v-v_r,r)\\
&=\phi_i(v_r,r).
\end{aligned}
\]
By (E) and nonnegativity,
$\sum_{i\in N}\phi_i(0,r)=0$ and hence $\phi_i(0,r)=0$ for every $i \in N$.
Consequently,
\[
(\forall i\in N)\qquad \phi_i(v,r)\le\phi_i(v_r,r).
\]
On the other hand, efficiency and $v_r(N,r)=v(N,r)$ give
\[
\sum_{i\in N}\bigl(\phi_i(v_r,r)-\phi_i(v,r)\bigr)
=v_r(N,r)-v(N,r)=0.
\]
Every summand is nonnegative, so each is zero. Therefore
\[
(\forall i\in N)\qquad \phi_i(v,r)=\phi_i(v_r,r).
\]
Since $v\in\mathcal V$ and $r\in X$ were arbitrary, this proves (2).
\end{proof}

The impossibility also applies to chronological credit
as a special case~\cite{FaigleKern1992}, since it satisfies (E), (A), and (N), and hence (SM).

\paragraph{Minimizing the credit of the creator.}
Assume a simpler setting where the total value $v(N,x)$ is independent of the construction $x$. In this case, society seeks to minimize the credit awarded to the current creator, while the creator seeks to maximize it. Theorem~\ref{thm:irrelevance}(2) implies that these objectives can be completely opposed.

\begin{corollary}\label{cor:fixed-welfare}
Suppose $|X|\ge2$. There exist $v\in\mathcal V$ and $b\in X$ with $v(N,x)=1$ for every $x\in X$ such that every rule $\phi$ satisfying \textnormal{(E)}, \textnormal{(SM)}, and \textnormal{(N)} satisfies
\begin{itemize}
	\item $\arg\max_{r\in X}\phi_0(v,r)=\{b\}$ and $\max_{r\in X}\phi_0(v,r) = 1$.
	\item $\arg\min_{r\in X}\phi_0(v,r)=X\setminus\{b\}$ and $\min_{r\in X}\phi_0(v,r) = 0$.
\end{itemize}
%
\end{corollary}
\begin{proof}
Fix $b\in X$ and define
\[
(\forall S\subseteq N)\,(\forall x\in X)\qquad
v(S,x)=\mathbf1_{\{x=b,\,0\in S\}}+\mathbf1_{\{x\ne b,\,1\in S\}}.
\]
This $v$ belongs to $\mathcal V$, and $v(N,x)=1$ for all $x\in X$.
By Theorem~\ref{thm:irrelevance}(2), every rule $\phi$ satisfying (E) and (SM) has
$\phi_0(v,r)=\phi_0(v_r,r)$ for every $r\in X$.
In $v_b$ every player $i\ne0$ is null; hence (N) and (E) imply
$\phi_0(v,b)=\phi_0(v_b,b)=v_b(N,b)=1$.
For each $r\ne b$, player $0$ is null in $v_r$, so (N) gives
$\phi_0(v,r)=\phi_0(v_r,r)=0$. Both claimed equalities follow.
\end{proof}

\paragraph{Without submodularity.}
Without (SM), axioms (E), (N), and (IC) can hold together,
(and naturally (A) which is stronger than (SM) must also be dropped). At a high level, this resembles
common acknowledgment practice, such as scientific citations or software
dependency lists: relevant prior work is acknowledged
without assigning numerical weights to individual contributions.

\begin{observation}
There exists a credit rule satisfying \textnormal{(E)}, \textnormal{(N)},
and \textnormal{(IC)}.
\end{observation}
\begin{proof}
Let $A(v)$ be the set of influential players for $v$:
\[
(\forall v\in\mathcal V)\qquad
A(v)=\bigl\{i\in N:\ (\exists x\in X)\,(\exists S\subseteq N\setminus\{i\})\ \ \Delta_i v(S,x)>0\bigr\}.
\]
Define
\[
\begin{aligned}
&(\forall v\in\mathcal V)\,(\forall r\in X)\,(\forall i\in N)\\
&\qquad\phi_i(v,r)=\begin{cases}
\dfrac{v(N,r)}{|A(v)|},&A(v)\ne\varnothing\text{ and }i\in A(v),\\
0,&\text{otherwise}.
\end{cases}
\end{aligned}
\]
Since $v(N,r)\ge0$, this defines a nonnegative credit rule.
Fix $v\in\mathcal V$ and $r\in X$. If $A(v)=\varnothing$, then
\[
(\forall i\in N)\,(\forall x\in X)\,(\forall S\subseteq N\setminus\{i\})\qquad
\Delta_i v(S,x)=0.
\]
Let $i_1,\ldots,i_{n+1}$ be an ordering
of $N$ and $S_k=\{i_1,\ldots,i_k\}$ for every $k\in\{1,\ldots,n+1\}$, with $S_0=\varnothing$. Thus
\[
v(N,r)=v(\varnothing,r)+
\sum_{k=1}^{n+1}\Delta_{i_k}v(S_{k-1},r)=0.
\]
Consequently, $\sum_{i\in N}\phi_i(v,r)=0=v(N,r)$.
If $A(v)\ne\varnothing$, then
\[
\sum_{i\in N}\phi_i(v,r)
=\sum_{i\in A(v)}\frac{v(N,r)}{|A(v)|}=v(N,r).
\]
Hence (E) holds in both cases. If the premise of (N) holds for a player $i\in N$, that is,
\[
(\forall x\in X)\,(\forall S\subseteq N\setminus\{i\})\qquad \Delta_i v(S,x)=0,
\]
then $i\notin A(v)$.
Therefore $\phi_i(v,r)=0$ for every $r\in X$, proving (N).
Finally, if player $0$ is influential, then $0\in A(v)$ and
$|A(v)|>0$. Hence
\[
(\forall r\in X)\qquad
\phi_0(v,r)=\frac{v(N,r)}{|A(v)|}.
\]
Since $|A(v)|>0$ is independent of $r$, we obtain
\[
\arg\max_{r\in X}\phi_0(v,r)=\arg\max_{r\in X}v(N,r).
\]
Thus, (IC) holds.
\end{proof}

Without (SM), uniqueness generally fails: for example, both equal division
of $v(N,r)$ among influential players and division proportional to
$1+\sum_{x\in X}v(\{i\},x)$ among those players satisfy (E), (N), and (IC).
Thus, additional axioms are needed to characterize a unique credit rule.

\paragraph{Uniqueness of Shapley in the Counterfactual world.}
The above observation does not prove uniqueness. We now show that using the additional axiom of symmetry, the classic uniqueness result of Shapley~\cite{Shapley1953} holds even when there are counterfactual alternatives. This credit assignment is of course counterfactual-independent by Theorem~\ref{thm:irrelevance}.
Consider the additional \emph{symmetry} axiom.
For every $v\in\mathcal V$, $r\in X$, and distinct $i,j\in N$,
symmetry requires
\begin{equation}\tag{S}\label{ax:S}
	\Bigl[(\forall x\in X)(\forall S\subseteq N\setminus\{i,j\})\,
	\Delta_i v(S,x)=\Delta_j v(S,x)\Bigr]
	\implies \phi_i(v,r)=\phi_j(v,r).
\end{equation}

The four axioms (E), (A), (N), and (S) are considered the fairness axioms of credit assignment~\cite{Shapley1953}.
Define the Shapley value of the realized game as
\[
\begin{aligned}
&(\forall v\in\mathcal V)\,(\forall r\in X)\,(\forall i\in N)
&\qquad\operatorname{Sh}_i(v(\cdot,r))
=\sum_{S\subseteq N\setminus\{i\}}
\frac{|S|!(|N|-|S|-1)!}{|N|!} \cdot \Delta_i v(S,r).
\end{aligned}
\]
We show that even with counterfactual alternatives, the Shapley value of the realized game is the unique fair credit allocation. Note that by Theorem~\ref{thm:irrelevance} this allocation is not (IC) when there are counterfactual results $|X| \geq 2$.
\begin{corollary}
The unique credit rule $\phi$ satisfying \eqref{ax:E}, \eqref{ax:A},
\eqref{ax:N}, and \eqref{ax:S} is given by
\[
(\forall v\in\mathcal V)\,(\forall r\in X)\,(\forall i\in N)\qquad
\phi_i(v,r)=\operatorname{Sh}_i(v(\cdot,r)).
\]
\end{corollary}

\begin{proof}
By Theorem~\ref{thm:irrelevance}(2) (trivially if $|X|=1$), it suffices to consider
the realized game. The uniqueness result of Shapley~\cite{Shapley1953} applies to nonnegative monotone games;
see~\cite[Section 2]{AumannShapley1974}.
\end{proof}

\paragraph{Funding.} IDA is supported by grant NSF DMS-2031883 and Vannevar Bush Faculty Fellowship ONR-N00014-20-1-2826 (PI Mossel). EM is partially supported by NSF DMS-2031883, Vannevar Bush Faculty Fellowship ONR-N00014-20-1-2826, MURI N000142412742, and a Simons Investigator Award.

\paragraph{AI Training.} The authors do not authorize AI training on this paper.

\newpage


\begin{thebibliography}{99}
\bibitem{Shapley1953}
L.~S.~Shapley.
A value for $n$-person games.
In H.~W.~Kuhn and A.~W.~Tucker, editors,
\emph{Contributions to the Theory of Games, Volume II},
pages 307--318. Princeton University Press, 1953.


\bibitem{ShenBarabasi2014}
H.-W.~Shen and A.-L.~Barabási.
Collective credit allocation in science.
\emph{Proceedings of the National Academy of Sciences},
111(34):12325--12330, 2014.

\bibitem{FaigleKern1992}
U.~Faigle and W.~Kern.
The Shapley value for cooperative games under precedence constraints.
\emph{International Journal of Game Theory}, 21:249--266, 1992.

\bibitem{Euler1740}
L.~Euler.
De summis serierum reciprocarum.
\emph{Commentarii Academiae Scientiarum Petropolitanae},
7:123--134, 1740.
Presented in 1735.

\bibitem{Wastlund2010}
J.~Wästlund.
Summing inverse squares by Euclidean geometry.
Chalmers University of Technology, 2010.
\url{https://www.math.chalmers.se/~wastlund/Cosmic.pdf}.

\bibitem{Weber1988}
R.~J.~Weber.
Probabilistic values for games.
In A.~E.~Roth, editor,
\emph{The Shapley Value: Essays in Honor of Lloyd S.~Shapley},
pages 101--120. Cambridge University Press, 1988.

\bibitem{Frye2020}
C.~Frye, C.~Rowat, and I.~Feige.
Asymmetric Shapley values: Incorporating causal knowledge into
model-agnostic explainability.
In \emph{Advances in Neural Information Processing Systems},
volume 33, 2020.

\bibitem{AumannShapley1974}
R.~J.~Aumann and L.~S.~Shapley.
\emph{Values of Non-Atomic Games}.
Princeton University Press, 1974.

\bibitem{Heskes2020}
T.~Heskes, E.~Sijben, I.~G.~Bucur, and T.~Claassen.
Causal Shapley values: Exploiting causal knowledge to explain
individual predictions of complex models.
In \emph{Advances in Neural Information Processing Systems},
volume 33, pages 4778--4789, 2020.

\bibitem{Foerster2018}
J.~Foerster, G.~Farquhar, T.~Afouras, N.~Nardelli, and S.~Whiteson.
Counterfactual multi-agent policy gradients.
In \emph{Proceedings of the AAAI Conference on Artificial Intelligence},
volume 32, pages 2974--2982, 2018.

\bibitem{Mesnard2021}
T.~Mesnard, T.~Weber, F.~Viola, S.~Thakoor, A.~Saade,
A.~Harutyunyan, W.~Dabney, T.~S.~Stepleton, N.~Heess,
A.~Guez, E.~Moulines, M.~Hutter, L.~Buesing, and R.~Munos.
Counterfactual credit assignment in model-free reinforcement learning.
In \emph{Proceedings of the 38th International Conference on
Machine Learning}, volume 139 of \emph{Proceedings of Machine
Learning Research}, pages 7654--7664, 2021.

\bibitem{Meulemans2023}
A.~Meulemans, S.~Schug, S.~Kobayashi, N.~Daw, and G.~Wayne.
Would I have gotten that reward? Long-term credit assignment
by counterfactual contribution analysis.
In \emph{Advances in Neural Information Processing Systems},
volume 36, 2023.

\bibitem{Li2021}
J.~Li, K.~Kuang, B.~Wang, F.~Liu, L.~Chen, F.~Wu, and J.~Xiao.
Shapley counterfactual credits for multi-agent reinforcement learning.
In \emph{Proceedings of the 27th ACM SIGKDD Conference on
Knowledge Discovery and Data Mining},
pages 934--942, 2021.

\bibitem{Li2026}
M.~Li, K.-Z.~Lee, and E.~Bareinboim.
Counterfactual Shapley credit assignment.
\emph{Reinforcement Learning Journal}, 7, 2026.

\bibitem{Kotthoff2018}
L.~Kotthoff, A.~Fréchette, T.~P.~Michalak, T.~Rahwan,
H.~H.~Hoos, and K.~Leyton-Brown.
Quantifying algorithmic improvements over time.
In \emph{Proceedings of the 27th International Joint Conference
on Artificial Intelligence (IJCAI)}, pages 5165--5171, 2018.

\bibitem{OzerturkYildirim2021}
S.~Ozerturk and H.~Yildirim.
Credit attribution and collaborative work.
\emph{Journal of Economic Theory}, 195:105264, 2021.

\bibitem{OzerturkYildirim2025}
S.~Ozerturk and H.~Yildirim.
Who gets the credit? Credit attribution, spillovers,
and inefficiency in teams.
\emph{Games and Economic Behavior}, 154:246--266, 2025.

\bibitem{Sharma2022}
A.~Sharma, H.~Li, and J.~Jiao.
The counterfactual-Shapley value: Attributing change in system metrics.
In \emph{NeurIPS 2022 Workshop on Causal Machine Learning
for Real-World Impact}, 2022.

\bibitem{Bilodeau2024}
B.~Bilodeau, N.~Jaques, P.~W.~Koh, and B.~Kim.
Impossibility theorems for feature attribution.
\emph{Proceedings of the National Academy of Sciences},
121(2):e2304406120, 2024.

\end{thebibliography}
\end{document}